\documentclass[fleqn,10pt]{wlpeerj}

\usepackage[ruled]{algorithm2e}

\SetAlFnt{\small}
\SetAlCapFnt{\small}
\SetAlCapNameFnt{\small}
\SetAlCapHSkip{0pt}
\IncMargin{-\parindent}

\usepackage{epsfig}
\usepackage{url}
\usepackage{listings}
\newcommand{\oh}[1]
    {\mbox{$ {\mathcal O}( #1 ) $}}

\newcommand{\fig}[1]
    {Figure~\ref{fig:#1}}

\newcommand{\figs}[2]
    {Figures~\ref{fig:#1} and~\ref{fig:#2}}

\title{QuickSched: Task-based parallelism with dependencies and conflicts}
\author[1,3]{Pedro Gonnet}
\author[1]{Aidan B. G. Chalk}
\author[2]{Matthieu Schaller}
\affil[1]{School of Engineering and Computing Sciences, Durham University, United Kingdom.}
\affil[2]{Institute for Computational Cosmology, Durham University, United Kingdom.}
\affil[3]{Google Switzerland GmbH, Z\"urich, Switzerland.}


\begin{abstract}
This paper describes QuickSched, a compact and efficient Open-Source
C-language library for task-based shared-memory parallel programming.
QuickSched extends the standard dependency-only scheme of task-based
programming with the concept of task conflicts, i.e.~sets of tasks
that can be executed in any order, yet not concurrently.
These conflicts are modelled using exclusively lockable
hierarchical resources.
The scheduler itself prioritizes tasks along the critical path
of execution and is shown to perform and scale well on a 64-core parallel
shared-memory machine for two example problems: A tiled QR
decomposition and a task-based Barnes-Hut tree code.
\end{abstract}

\begin{document}

\flushbottom
\maketitle
\thispagestyle{empty}

\section{Introduction}

Task-based parallelism is a conceptually simple paradigm for
shared-memory parallelism in which a computation is broken-down
into a set of inter-dependent tasks which are executed
concurrently.
Task dependencies are used to model the flow of data between
tasks, e.g.~if task $B$ requires some data generated by task $A$,
then task $B$ {\em depends} on task $A$ and cannot be executed
before task $A$ has completed.
The tasks and their dependencies can be seen as the nodes and edges,
respectively, of a Directed Acyclic Graph (DAG) which can be
traversed in topological order, executing the tasks at the nodes
on the way down.

Once modelled in such a way, the computation is somewhat trivial
to parallelize:
given a set of inter-dependent tasks and a set of computational
threads, each thread repeatedly selects a task with no
unsatisfied dependencies from the DAG and executes it.
If no tasks are available, the thread waits until any other
thread finishes executing a task, thus potentially releasing
new tasks, or until all tasks in the DAG have been executed.
Note that although the parallel execution
itself is trivial, it does not always guaranteed to be efficient.
Several factors may limit the maximum degree of parallelism, e.g.~the
structure of the task dependency DAG itself, or the order in which
available tasks are executed.

\begin{figure}
    \centerline{\epsfig{file=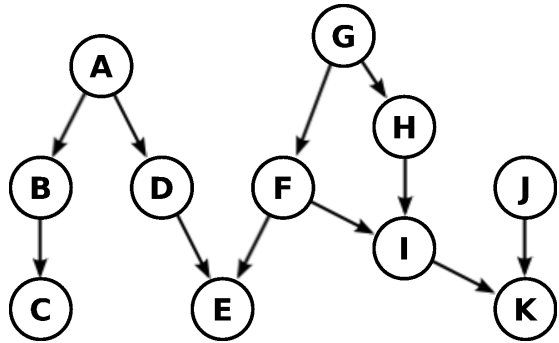,width=0.5\textwidth}}
    \caption{A set of tasks (circles) and their dependencies (arrows).
        The arrows indicate the direction of the dependency, i.e.~an
        arrow from task $A$ to task $B$ indicates that task $B$ depends
        on task $A$.
        Tasks $A$, $G$, and $J$ have no unsatisfied dependencies and
        can therefore be executed.
        Once task $G$ has completed, tasks $F$ and $H$ become available,
        and task $E$ only becomes available once both tasks $D$ and $F$
        have completed.}
    \label{fig:Tasks}
\end{figure}

\fig{Tasks} shows such a DAG for a set of tasks with
arrows indicating the direction of the dependencies, i.e.~an
arrow from task $A$ to task $B$ indicates that task $B$ depends
on task $A$.
In a parallel setting, tasks $A$, $G$, and $J$ can be
executed concurrently.
Once task $G$ has completed, tasks $F$ and $H$ become available
and can be executed by any other computational thread.

One of the first implementations of a task-based parallel programming
systems is Cilk \citep{ref:Blumofe1995}, an extension to the C
programming language which allows function calls to be ``spawned''
as new tasks.
Dependencies are enforced by the {\tt sync} keyword, which
forces a thread to wait for all the tasks that it spawned
to complete.

In SMP superscalar \citep{ref:Perez2008}, StarPU \citep{ref:Augonnet2011},
QUARK \citep{ref:Yarkhan2011}, and KAAPI \citep{ref:Gautier2007}
the programmer specifies
what shared data each task will access, and how that data will
be accessed, e.g.~read, write, or read-write access.
The dependencies between tasks are then generated
automatically by the runtime system, assuming that the
data must be accessed and updated in the order in which
the tasks are generated.
StarPU also provides an interface for specifying additional
dependencies explicitly.
Intel's Threading Building Blocks (TBB)
\citep{ref:Reinders2010}
provide task-based parallelism using C++ templates in which
dependencies are handled either by explicitly waiting
for spawned tasks, or by explicitly manipulating
task reference counters.

Finally, the very popular OpenMP standard provides some basic support
for spawning tasks, similar to Cilk, as of version 3.0
\citep{ref:OpenMP2008}.
OmpSs \citep{ref:Duran2011} extends this scheme with automatic
dependency generation as in SMP superscalar, of which it
is a direct descendant, along with
the ability to explicitly wait on certain tasks.

\begin{figure}
    \centerline{\epsfig{file=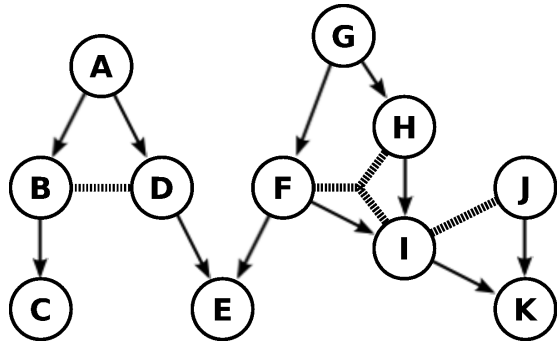,width=0.5\textwidth}}
    \caption{Task graph with conflicts (thick dashed lines).
        If two or more tasks are joined by a conflict, they cannot be
        executed concurrently, i.e. tasks $B$ and $D$ cannot be run an
        the same time.
        Tasks belonging to different conflicting sets, e.g. tasks $B$
        and $F$, or tasks $F$ and $J$, however, can be executed
        concurrently.}
    \label{fig:TaskConflicts}
\end{figure}

In all of these systems, the tasks are only aware of a single
type of relationship between each other, i.e. dependencies, which
specify a strict ordering between two tasks.
In many cases, however, the task ordering need not necessarily
be this strict.
Consider the case of two tasks that update some shared resource
in an order-independent way, e.g. when accumulating a result in
a shared variable, or exclusively writing to an output file.
In order to avoid concurrent access to that resource, it is
imperative that the execution of both tasks does not overlap,
yet the order in which the tasks are executed is irrelevant.
In the following, such a relationship will be referred to
as a {\em conflict} between two tasks.
\fig{TaskConflicts} shows a task graph extended by conflicting tasks
joined by thick dashed lines.
None of tasks $F$, $H$, and $I$ can be executed concurrently,
i.e. they must be serialized, yet in no particular order.

In dependency-only systems, such conflicts can be modelled
with dependencies, which enforce a pre-determined arbitrary
ordering on conflicting tasks.
This artificial restriction on the order
in which tasks can be scheduled can, however, severely limit the
parallelizability of a computation, especially in the presence
of multiple conflicts per task.
Both \cite{ref:Ltaief2012} and \cite{ref:Agullo2013} note
this problem in their respective implementations of the Fast Multipole
Method (FMM), in which forces computed in different tasks are
accumulated on a set of particles.

Several libraries provide some mechanism to model such
conflicts, either directly or indirectly.
In the QUARK scheduler, conflicts can be modeled by explicitly
marking dependencies as concurrent.
KAAPI and OmpSS, on the other hand, allow marking access to
certain variables as reductions, yet only for basic operations,
e.g.~summation or maximum/minimum.

This paper presents QuickSched, a framework for task-based
parallel programming with constraints, which aims to achieve
the following goals:
\begin{itemize}
    \item {\em Correctness}: All constraints, i.e.~dependencies and
        conflicts, must be correctly enforced,
    \item {\em Speed}: The overheads associated with task management
        should be as small as possible,
    \item {\em Memory/cache efficiency}: Tasks accessing similar
        sets of data should be preferentially executed on the
        same core to preserve memory/cache locality as far as possible, and
    \item {\em Parallel efficiency}: The order in which the tasks
        are executed should be chosen such
        that sufficient work is available for all computational
        threads at all times.
\end{itemize}
\noindent 
Section~2 describes the main design considerations, Section~3 the
underlying algorithms and data structures, and
Section~4 their specific implementation in QuickSched.
Section~5 presents two test-cases:
\begin{enumerate}
    \item The tiled QR
    decomposition described in \cite{ref:Buttari2009} and for
    which the QUARK scheduler was originally developed, and
    \item A task-based Barnes-Hut tree-code to compute the
    gravitational N-body problem similar to the FMM codes
    of \cite{ref:Ltaief2012} and \cite{ref:Agullo2013},
\end{enumerate}
These real-world examples show how QuickSched can be used in practice,
and can be used to assess its efficiency.
Section~6 concludes with some general observations and future work
directions.

\section{Design Considerations}

From a programmer's perspective, there are two main paradigms for generating
task dependencies:
\begin{itemize}
  \item Implicitly via spawning and waiting, e.g.~as is done in Cilk
    and OpenMP~3.0, or
  \item Automatic extraction from data dependencies, e.g.~as is done in
    StarPU, QUARK, and OmpSs.
\end{itemize}

\begin{figure}
    \centerline{\epsfig{file=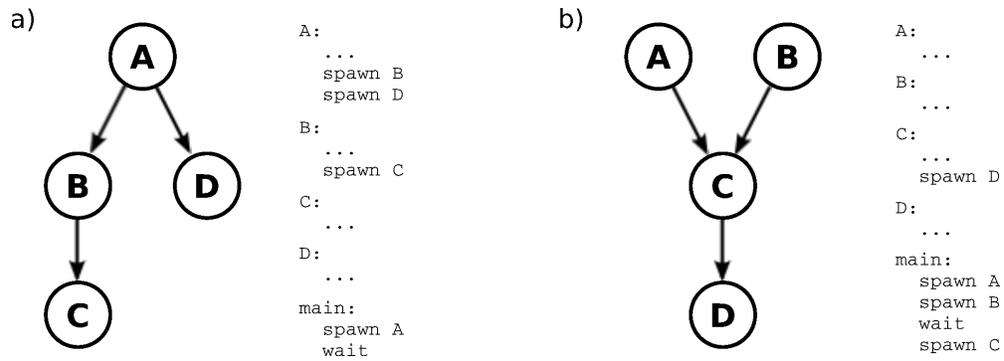,width=0.9\textwidth}}
    \caption{Two different task graphs and how they can be implemented
      using spawning and waiting.
      For the task graph on the left, each task spawns its dependent
      task or tasks. For the task graph on the right, in which task $C$
      has multiple dependencies, tasks $A$ and $B$ must be spawned and waited
      for by the calling thread before task $C$ can be spawned.}
    \label{fig:Spawn}
\end{figure}

The first scheme, spawning and waiting, is arguably the simplest to
use.
For simple depedency structures in which each task depends on only a
single task, i.e.~if the task DAG is a tree, each task {\em spawns}, or
creates, its dependent tasks after its completion (see \fig{Spawn}a).
Hence for the tasks $A$--$E$ in \fig{Tasks}, task $A$ would spawn
tasks $B$ and $D$, task $B$ would spawn task $C$, and task $D$ would
spawn task $E$.
If a task has more than one dependency, e.g. tasks $D$--$F$ in \fig{Tasks},
then the task generation order is reversed: Task $E$ is executed first,
and first spawns tasks $D$ and $F$, and waits for both their completion
before doing its own computations (see \fig{Spawn}b).

Although simple to use, this implicit dependency management
limits the types of DAGs that can be represented, e.g.~for
all the tasks in \fig{Tasks}, using such a spawning and waiting model
would create implicit dependencies between the lowest-level
tasks $C$, $E$, and $K$.
The main thread would spawn tasks $A$, $G$ and $J$, $A$ spawns $B$ and $D$,
$G$ spawns $F$, $H$, and then $I$, $B$ spawns $C$.
The main thread then has to wait for $A$, $G$, and $J$,
and thus implicitly all their spawned tasks, before executing
$E$ and $K$.

Automatic dependency extraction, on the other hand,
works by enforcing dependencies
between tasks that access the same data.
These data dependencies are provided explicitly by the programmer, e.g.~by
describing which parameters to a task are input, output, and input/output.
The dependencies are enforced in the order in which the tasks are 
created.
This approach usually relies on compiler extensions, e.g.~{\tt pragma}s
in C/C++, or a system of function call wrappers, to describe the task parameters
and their intent.

This approach allows programmers to specify rich dependency hierarchies
with very little effort, i.e.~without having to explicitly think about
dependencies at all, yet they still only allow for one type of relationship,
i.e.~dependencies, and lack the ability to deal with conflicts as
described in the previous section.
They may also not be able to understand more complex memory access patterns,
e.g.~for two tasks modifying the upper and lower triangular parts of a matrix,
which access the same block of memory, but never actually generate any
concurrency issues.

Here, QuickSched takes a drastically different approach by requiring
the programmer to create the complete task graph and its dependencies
explicitly before execution.
This approach has two main advantages:
\begin{itemize}
  \item It gives the user maximum flexibility with regards to the
    structure of the task graph generated,
  \item Knowing the complete structure of the task graph before execution
    allows the task scheduler to make more informed decisions
    as to how the tasks should be prioritized.
\end{itemize}

\noindent The obvious disadvantage is the burden of producing a correct
task graph is placed on the programmer.
Although some problems such as cyclic dependencies can be detected
automatically, there is no automatic way to detect whether the
dependencies actualy reflect the intent of the programmer.
Due to this added complexity, we consider QuickSched to be
a tool not designed for casual parallel programmers, but for 
those interested in investing a bit more programming effort to achieve
better performance.

As opposed to the dependencies,
conflicts between tasks or groups of tasks are not specified directly,
but are instead modeled as exclusive locks on a shared resource
which have to be obtained by a task before it can execute.
Thus, in \fig{TaskConflicts}, before executing, task $F$ has
to obtain an exclusive lock on the resource associated with
the conflict between tasks $F$, $H$, and $I$.
While task $F$ is being executed, neither $H$ nor $I$ can 
lock the same resource, and therefore will not execute until
task $F$ is done and the lock has been released.

As with all other task-based libraries, the partitioning of the
computation into tasks is also left entirely to the programmer.
In theory, any program can be automatically converted to a task-based
representation since each statement in the program code
can be considered a single task, with dependencies to the
statements/tasks that produce the input values.
This set of basic tasks could, again in theory, be reduced by merging
tasks that share dependencies and/or resources using a variety
of graph algorithms.
The decomposition of a computation into tasks, however, usually
involves re-thinking the underlying algorithms such that they
best fit the task-based paradigm, e.g.~as in the examples in the
following sections, or as in \cite{ref:Gonnet2014,ref:Buttari2009,ref:Ltaief2012}.
This process requires careful evaluation of the underlying
computation, and is probably best
not left as an automatic transformation of an otherwise serial code.

Finally, the task granularity is an important issue: if the task
decomposition is too coarse, then good parallelism
and load-balancing will be difficult to achieve.
Converseley, if the tasks are too small, the costs of selecting and
scheduling tasks, which is usually constant per task, will
quickly destroy any performance gains from parallelism.
Starting from a per-statement set of tasks, it is therefore
reasonable to group them by their dependencies and shared resources.

In the examples presented herein, we have chosen our task decomposition
and granularity such that
\begin{itemize}
  \item Each task maximizes the ratio of computation to data required,
  \item The resources required for each task fit comfortably in the
    lowest-level caches of the underlying system.
\end{itemize}
\noindent The first critera is biased towards bigger tasks, while the
second limits their size.
The parameters controlling the size of the tasks in the examples, 
i.e.~the tile size in the QR decomposition and the limits $n_\mathsf{max}$
and $n_\mathsf{task}$ were determined empirically and only optimized
to the closest power of two or rough power of ten, respectively.
Further tuning these parameters could very likely lead to further
performance gains, but such an effort would go beyond the scope,
and point, of this paper.

\section{Data Structures and Algorithms}

The QuickSched task scheduler consists of four main
objects types: {\em task}, {\em resource}, {\em scheduler},
and {\em queue}.
The task and resource objects are used
to model the computation, i.e. the work that is to be done
and the data on which it will be done, respectively.
The scheduler and queue objects manage
how the work is done, i.e. which tasks get scheduled
where and when, respectively.

\begin{figure}
    \centerline{\epsfig{file=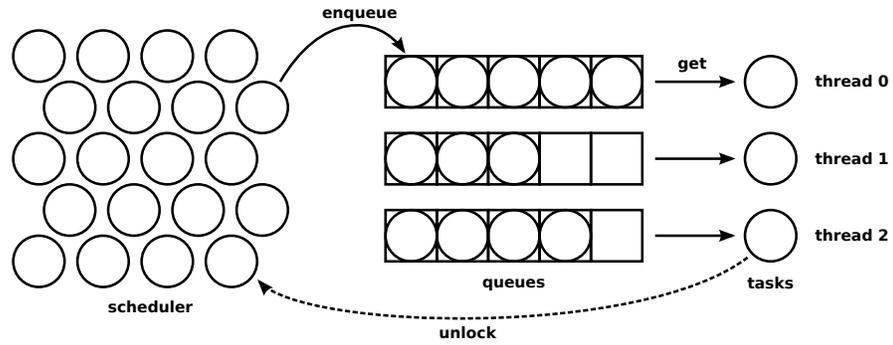,width=0.8\textwidth}}
    \caption{Schematic of the QuickSched task scheduler.
        The tasks (circles) are stored in the scheduler (left).
        Once a task's dependencies have been resolved, the task
        is moved to one of the task queues.
        Tasks that are not involved in any active conflicts
        are then taken from the queues by the different
        computational threads and executed.
        After execution, their dependent tasks are unlocked
        in the scheduler (dashed arrow).}
    \label{fig:QSched}
\end{figure}

The division of labor between the scheduler and
the queue objects is illustrated in \fig{QSched}.
The scheduler holds the tasks and is in charge
of managing {\em dependencies}.
Once a task has no unresolved dependencies, it is passed
on to a queue object.
The queue object, on the other hand, is in charge
of managing {\em conflicts}.
Computational threads can query a queue and will
receive only tasks for which all conflicts have been
resolved, i.e. for which all necessary resources could be 
exclusively locked.

There is also a division of responsibilities regarding
{\em efficiency} between the scheduler and the queue
objects.
The tasks in each queue are grouped according to the resources
they use, i.e. all the tasks in the same queue use a
similar set of resources.
The underlying assumption is that each computational
thread will preferentially access the same queue for tasks.
If the tasks in the queue share the same set of resources,
it increases the probability of said resources already
being present in the thread's cache, thus increasing
{\em cache efficiency}.
The scheduler is in charge of selecting the most appropriate
queue for each task, based on information stored in each task
on which resources are used.
Given a set of tasks with similar resources for which all
dependencies are resolved, it is up to the queue to decide which
tasks to prioritize.

The following subsections describe these four object types
in detail, as well as their operations.

\subsection{Tasks}
\label{sec:tasks}

A task consists of the following data structure, in C-like pseudo-code:

\begin{center}\begin{minipage}{0.9\textwidth}
    \begin{lstlisting}
struct task {
  int type, wait;
  void *data;
  struct task **unlocks;
  struct resource **locks, **uses;
  int size_data, nr_unlocks, nr_locks, nr_uses;
  int cost, weight;
};
    \end{lstlisting}
\end{minipage}\end{center}
\noindent where the {\tt data}, {\tt unlocks}, {\tt locks},
and {\tt uses} arrays are pointers to the contents of other
arrays, i.e.~they are not allocated individually.

{\em What} the task does is determined by the {\tt type}
field, e.g.~which can be mapped to any particular function,
and the {\tt data} pointer which points to an array of
{\tt size\_data} bytes containing data specific to the task,
e.g.~the parameters for a specific function call.
Both fields are application-specific and therefore not
important for the scheduler itself.

The {\tt unlocks} field points to the first element of
an array of {\tt nr\_unlocks} pointers to other tasks.
These pointers represent the dependencies in reverse:
if task $B$ depends on task $A$, then task $A$ {\em unlocks}
task $B$.
The unlocks therefore follow the direction of the arrows
in \figs{Tasks}{TaskConflicts}.
Conversely, {\tt wait} is the number of unresolved dependencies
associated with this task, i.e.~the number of unexecuted tasks
that unlock this task.
The wait-counters can
be set by initializing all the task wait counters to zero and then
incrementing the wait counter of each unlock-task of each task.

The {\tt locks} field of each task points to the first element of
an array of {\tt nr\_locks} pointers to {\em resources}
for which exclusive locks must be obtained for the task
to execute.
Each locked resource represents a task conflict.
Similarly, {\tt uses} points to the first element of
an array of {\tt nr\_uses} pointers to resources which
will be used, but need not be locked.

\begin{figure}
    \centerline{\epsfig{file=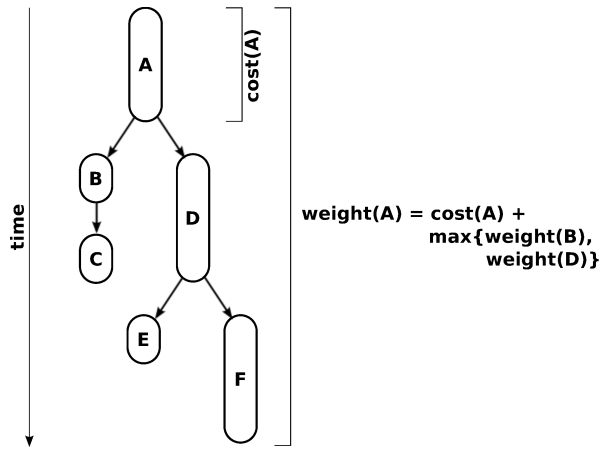,height=0.4\textwidth}}
    \caption{Computation of the task weight.
      In this task graph, the height of each task corresponds to its
      computational {\em cost}.
      The {\em weight} of each task, e.g.~the topmost task,
      is computed from its cost plus
      the maximum dependent task's weight.
      This corresponds to the computatoinal cost of the critical
      path following the task's dependencies.}
    \label{fig:TaskWeight}
\end{figure}

Finally, {\tt cost} and {\tt weight} are measures
for the relative computational cost of this task, and the
relative cost of the critical path following the
dependencies of this task, respectively, i.e.~the task's cost
plus the maximum dependent task's weight (see \fig{TaskWeight}).
The task cost can be either a rough estimate provided by the user,
or the actual cost of the same task last time it was executed.
The task weights are computed by traversing
the tasks DAG in reverse topological order following their dependencies,
e.g.~as per \cite{ref:Kahn1962} in $\mathcal O(n)$ for $n$ tasks,
and computing each task's weight, e.g.
\begin{equation*}
  \mbox{weight}_i = \mbox{cost}_i + \max_{j \in \mbox{\small unlocks}_i}\left\{\mbox{weight}_j\right\}.
\end{equation*}
\noindent where $\mbox{weight}_i$ and $\mbox{cost}_i$ are the
weight and cost of the $i$th task, respectively, and 
$\mbox{unlocks}_i$ is the set of tasks that the $i$th task
unlocks.

\subsection{Resources}

Resources consist of the following data structure:
\begin{center}\begin{minipage}{0.9\textwidth}
    \begin{lstlisting}
struct resource {
  struct resource *parent;
  volatile int lock, hold;
  int owner;
};
    \end{lstlisting}
\end{minipage}\end{center}

The {\tt parent} field, which points to another resource, is
used to create hierarchical resources, i.e.~resources
that are themselves subsets of larger resources.
This can be useful, e.g.~in the context of particle simulations
described in the next section, where particles are sorted
into hierarchical cells which are used at different levels.
The {\tt owner} field is the ID of the queue to which this
resource has been preferentially assigned.

\begin{figure}
    \centerline{\epsfig{file=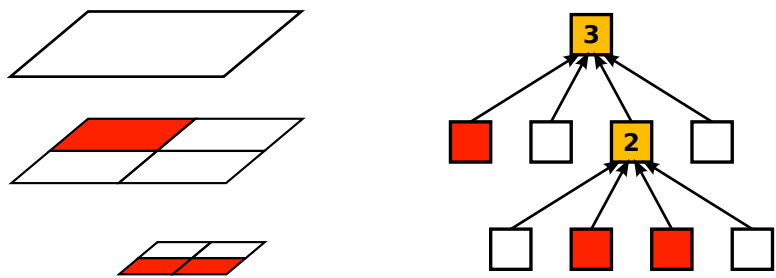,width=0.7\textwidth}}
    \caption{A hierarchy of cells (left) and the hierarchy of
        corresponding hierarchical resources at each level.
        Each square on the right represents a single resource, and
        arrows indicate the resource's parent.
        Resources colored red are locked, resources colored orange
        are held, where the number in the square indicates the
        value of the hold counter.}
    \label{fig:Resources}
\end{figure}

The {\tt lock} field is either {\tt 0} or {\tt 1} and indicates
whether this resource is currently in use, i.e.~{\em locked}.
To avoid race conditions, this value should be tested
and set using atomic instructions only.
The {\tt hold} field is a counter indicating how many
sub-resources of the current resource are locked.
If a resource's hold counter is not zero, then it is
{\em held} and cannot be locked.
Likewise, if a resource is locked, it cannot be held
(see \fig{Resources}).

Incrementing the hold counter of the resource can be implemented
as follows:
\begin{center}\begin{minipage}{0.9\textwidth}
    \begin{lstlisting}
void resource_hold(struct resource *r) {
  if (atomic_cas(&r->lock, 0, 1) != 0) return 0;
  atomic_inc(&r->hold);
  r->lock = 0;
  return 1;
}
    \end{lstlisting}
\end{minipage}\end{center}
\noindent where {\tt atomic\_cas(val,old,new)} is an atomic
compare-and-swap operation that sets {\tt val} to {\tt new}
if it is currently equal to {\tt old}.
Similarly, {\tt atomic\_inc(val)} increments {\tt val} by one
atomically.
The resource's {\tt lock} is used to check if the resource
is already locked (line 2), and it is held while {\tt hold}
is incremented, to avoid overlapping hold/lock operations.
If the resource can be locked, the hold counter is incremented
atomically (line~3), and the lock is released (line~4),
returning {\tt 1} or {\tt 0} if the resource could be held
or not, respectively.

The locking procedure itself is implemented as follows:
\begin{center}\begin{minipage}{0.9\textwidth}
    \begin{lstlisting}
void resource_lock(struct resource *r) {
  struct resource *up, *top;
  if (r->hold && atomic_cas(&r->lock, 0, 1) != 0)
    return 0;
  if (r->hold) {
    r->lock = 0;
    return 0;
  }
  for (up = r->parent; up != NULL; up = up->parent)
    if (!resource_hold(up)) break;
  if ((top = up) != NULL) {
    for (up = r->parent; up != top, up = up->parent)
      atomic_dec(&up->hold);
    r->lock = 0;
    return 0;
  } else
    return 1;
}
    \end{lstlisting}
\end{minipage}\end{center}
\noindent where in line~3 the resource is first locked if it
is not held.
Due to a possible race condition when holding the resource,
the {\tt hold} counter must be checked again once the resource
has been locked (line~5).
In lines~9--10 the hold counters of the hierarchical parents
are incremented using the procedure described earlier.
If this process fails at any point (line~11), the
previously set hold counters are decremented (line~13)
and the lock is released (line~14).
The procedure then returns {\tt 1} or {\tt 0} if the resource
could be locked or not, respectively.

Finally, unlocking a resource is relatively straight-forward,
i.e.~the lock is set to zero and the hold counters up the
resource hierarchy are decremented.

\subsection{Queues}

The main job of the task queues is, given a set of ready tasks,
to find the task with maximum weight, i.e.~the task along the
longest critical path, whose resources can all
be locked, and to do so as efficiently as possible.

One possible strategy would be to maintain an array of tasks
sorted by their weights, and to traverse that list in descending
order, trying to lock the resources of each task, until
a lockable task is found, or returning a failure otherwise.
Although this would return the best possible task, it
requires maintaining a sorted array in which inserting
or removing an entry is in \oh{n} for $n$ elements.
Using an unsorted array would require only \oh{1} operations for
insertion and deletion, but is undesirable as it completely
ignores the task weights.

As a compromise, the queue stores the tasks in an array
organized as a max-heap, i.e.~where the $k$th entry is ``larger''
than both the $2k+1$st and the $2k+2$nd entry,
with the task with maximum weight
in the first position.
Maintaining this heap structure requires \oh{\log n}
operations for both insertion and deletion, i.e. for the
bubble-up and trickle-down operations respectively.

Unfortunately, there is no way of efficiently traversing all
the elements in such a heap in decreasing order.
The array of tasks is therefore traversed as if it were sorted,
returning the first task that can be locked.
Although the first task in the array will be the task with
maximum weight, the following tasks are only loosely ordered,
where the $k$th of $n$ tasks has a larger weight than at least
$\lfloor n/k\rfloor -1$ other tasks.
Although this is not a particularly good lower bound, it turns
out to be quite sufficient in practice.

The data structure for the queue is defined as follows:
\begin{center}\begin{minipage}{0.9\textwidth}
    \begin{lstlisting}
struct queue {
  struct task **tasks;
  int count, lock;
};
    \end{lstlisting}
\end{minipage}\end{center}
\noindent where {\tt tasks} is an array of {\tt count} pointers
to the tasks in max-heap order, and {\tt lock} is used to
guarantee exclusive access to the queue.

Inserting a task in the queue is relatively straight-forward:
\begin{center}\begin{minipage}{0.9\textwidth}
    \begin{lstlisting}
void queue_put(struct queue *q, struct task *t) {
  while (atomic_cas(q->lock, 0, 1) != 0) {}
  q->tasks[q->count++] = t;
  bubble-up the q->count - 1st entry of q->tasks.
  q->lock = 0;
}
    \end{lstlisting}
\end{minipage}\end{center}
\noindent where the loop in line~2 spins until an exclusive
lock on the queue can be obtained.
The task is added to the end of the heap array (line~3)
and the heap order is fixed (line~4).
Before returning, the lock on the queue is released (line~5).

Obtaining a task from the queue can be implemented as follows:
\begin{center}\begin{minipage}{0.9\textwidth}
    \begin{lstlisting}
struct task *queue_get(struct queue *q) {
  struct task *res = NULL;
  int j, k;
  while (atomic_cas(q->lock, 0, 1) != 0) {}
  for (k = 0; k < q->count; k++) {
    for (j = 0; j < q->tasks[k]->nr_locks; j++)
      if (!resource_lock(q->tasks[k]->lock[j])) break;
    if (j < q->tasks[k]->nr_locks)
      for (j = j - 1; j >= 0; j--)
        resource_unlock(q->tasks[k]->lock[j]);
    else
      break;
  }
  if (k < q->count) {
    res = q->tasks[k];
    q->tasks[k] = q->tasks[--q->count];
    trickle-down the kth entry of q->tasks.
  }
  q->lock = 0;
  return res;
}
    \end{lstlisting}
\end{minipage}\end{center}
\noindent where, as with the queue insertion, the queue is first
locked for exclusive access (line~4).
The array of task pointers is then traversed (line~5),
locking the resources of each task (lines~6--7).
If any of these locks fail (line~8), the locks that were obtained
are released (lines~9--10), otherwise, the traversal is aborted
(line~12).
If all the locks on a task could be obtained (line~14), the
task pointer is replaced by the last pointer in the heap (line~16)
and the heap order is restored (line~17).
Finally, the queue lock is released (line~19) and the locked task,
or {\tt NULL} if no lockable task could be found, is returned.

Note that this approach of sequentially locking multiple resources
is prone to the so-called ``dining philosophers'' problem, i.e.~if
two tasks attempt, simultaneously, to lock the resources $A$ and $B$;
and $B$ and $A$, respectively, via separate queues, their respective calls
to {\tt queue\_get} will potentially fail perpetually.
This type of deadlock, however, is easily avoided by sorting the
resources in each task according to some global criteria, e.g.~the
resource ID or the address in memory of the resource.

Note also that protecting the entire queue with a mutex
is not particularly scalable, and several authors, e.g.~\cite{ref:Sundell2003},
have presented concurrent data structures that avoid this type
of locking.
However, since we normally use one queue per computational thread,
contention will only happens due to work-stealing, i.e.~when
another idle computational thread tries to poach tasks.
Since this happens only rarely, we opt for the simpler locking approach.
This decision is backed by the results in Section~5.

\subsection{Scheduler}

The scheduler object is used as the main interface to the
QuickSched task scheduler, and as such contains the instances
of the other three object types:
\begin{center}\begin{minipage}{0.9\textwidth}
    \begin{lstlisting}
struct qsched {
  struct task *tasks;
  struct queue *queues;
  struct resource *res;
  int nr_tasks, nr_queues, nr_resources;
  volatile int waiting;
};
    \end{lstlisting}
\end{minipage}\end{center}
\noindent where the only additional field {\tt waiting} is
used to keep track of the number of tasks that have not been
executed.
Note that for brevity, and to avoid conflicts with the naming
schemes of other standard libraries, the type name {\tt qsched}
is used for the scheduler data type.

The tasks are executed as follows:
\begin{center}\begin{minipage}{0.9\textwidth}
    \begin{lstlisting}
void qsched_run(qsched *s, void (*fun)(int, void *)) {
  qsched_start(s);
  #pragma omp parallel
  {
    int qid = omp_get_thread_num() % s->nr_queues;
    struct task *t;
    while ((t = qsched_gettask(s, qid)) != NULL) {
      fun(t->type, t->data);
      qsched_done(s, t);
    }
  }
}
    \end{lstlisting}
\end{minipage}\end{center}
\noindent where {\tt qsched\_start} initializes the tasks and
fills the queues (line~1).
For simplicity, OpenMP \citep{ref:Dagum1998}, which is available
for most compilers, is used to create a parallel section
in which the code between lines~4 and~11 is executed
concurrently.
A version using {\tt pthreads} \citep{ref:Pthreads1995}
directly\footnote{In most environments, OpenMP is implemented
on top of {\tt pthreads}, e.g. the {\tt gcc} compiler's libgomp.}
is also available.
The parallel section consists of a loop (lines~7--10) in
which a task is acquired via {\tt qsched\_gettask}
and its type and data are passed to a user-supplied
{\em execution function} {\tt fun}.
Once the task has been executed, it is returned to the
scheduler via the function {\tt qsched\_done}, i.e.~to
unlock its resources and unlock dependent tasks.
The loop terminates when the scheduler runs out of tasks,
i.e.~when {\tt qsched\_gettask} returns {\tt NULL}, and
the function exits once all the threads have exited their
loops.

At the start of a parallel computation, {\tt qsched\_start}
identifies the tasks that have no dependencies and sends them
to queues via the function {\tt qsched\_enqueue} which
tries to identify the best
queue for a given task by looking at which queues last used
the resources used and locked by the task, e.g.:
\begin{center}\begin{minipage}{0.9\textwidth}
    \begin{lstlisting}
void qsched_enqueue(qsched *s, struct task *t) {
  int best = 0, score[s->nr_queues];
  for (int k = 0; k < s->nr_queues; k++)
    score[k] = 0;
  for (int k = 0; k < t->nr_locks; k++) {
    int qid = t->locks[k]->owner;
    if (++score[qid] > score[best]) best = qid;
  }
  for (int k = 0; k < t->nr_uses; k++) {
    int qid = t->uses[k]->owner;
    if (++score[qid] > score[best]) best = qid;
  }
  queue_put(&s->queues[best], t);
}
    \end{lstlisting}
\end{minipage}\end{center}
\noindent where the array {\tt score} keeps a count of the
task resources ``owned'', or last used, by each queue.
The task is then sent to the queue with the highest such score
(line~13).

The function {\tt qsched\_gettask} fetches a task from
one of the queues:
\begin{center}\begin{minipage}{0.9\textwidth}
    \begin{lstlisting}
struct task *qsched_gettask(qsched *s, int qid) {
  struct task *res = NULL;
  int k;
  while (s->waiting) {
    if ((res = queue_get(s->queues[qid])) == NULL) {
      loop over all other queues in random order with index k
        if ((res = queue_get(s->queues[k])) != NULL)
          break;
    } else
      break;
  }
  if (res != NULL && s->reown) {
    for (k = 0; k < res->nr_locks; k++)
      res->locks[k]->owner = qid;
    for (k = 0; k < res->nr_uses; k++)
      res->uses[k]->owner = qid;
  }
  return res;
}
    \end{lstlisting}
\end{minipage}\end{center}
\noindent where the parameter {\tt qid} is the index of the
preferred queue.
If the queue is empty, or all of the tasks in that queue had
unresolved conflicts, the scheduler uses a variant of the
{\em work stealing} described in \cite{ref:Blumofe1999},
i.e.~it loops over all other queues
in a random order (line~6) and tries to get a task from them
(line~7).
If a task could be obtained from any queue and task re-owning
is enabled (line~12),
the resources it locks and uses are marked as now being owned
by the preferred queue (lines~13--16).
Finally, the task, or {\tt NULL} if no task could be obtained,
is returned.

The final step in a task's life cycle is, on completion,
to unlock the resources and tasks which depend on it.
This is handled by the function {\tt qsched\_done}, which
calls {\tt qsched\_enqueue} on any tasks for which the
wait counter is decremented to zero.
Once all the dependent tasks have been unlocked, the
{\tt waiting} counter of the scheduler is decremented.

\section{Validation}

This section presents two test cases that show
how QuickSched can be used in real-world applications, and
provides benchmarks to assess its efficiency and scalability.
The first test is the tiled QR decomposition originally
described in \cite{ref:Buttari2009}, which has been used as a benchmark
by other authors \citep{ref:Agullo2009b,ref:Badia2009,ref:Bosilca2012}.
This example only requires dependencies and is presented 
as a point of comparison to existing task-based parallel
programming infrastructures.
The second example is a Barnes-Hut tree-code, a problem
similar to the Fast Multipole Method described in both
\cite{ref:Ltaief2012} and \cite{ref:Agullo2013}.
This example shows how conflicts, modeled
via hierarchical resources, can be useful in modelling and executing
a problem efficiently.

The source code of both examples is distributed with the
QuickSched library, along with scripts to run the benchmarks
and generate the plots used in the following.
All examples were compiled with gcc v.\,5.2.0 using the
{\tt -O2 -march=native} flags and run on
a 64-core AMD Opteron 6376 machine at 2.67\,GHz.

\subsection{Task-Based QR Decomposition}

\cite{ref:Buttari2009} introduced the concept of using task-based
parallelism for tile-based algorithms in numerical linear algebra,
presenting parallel codes for the Cholesky, LU, and QR
decompositions.
These algorithms are now part of the PLASMA and MAGMA
libraries for parallel linear algebra \citep{ref:Agullo2009}.
The former uses the QUARK task scheduler, which was originally
designed for this specific task, while the latter currently uses
the StarPU task scheduler \citep{ref:Agullo2011}.

\begin{figure}
    \centerline{\epsfig{file=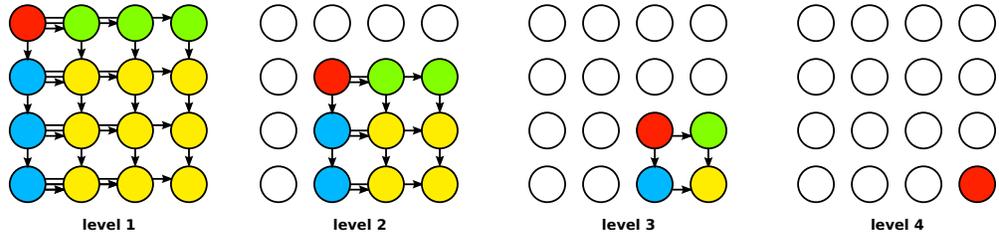,width=0.9\textwidth}}
    \caption{Task-based QR decomposition of a matrix consisting
        of $4\times 4$ tiles.
        Each circle represents a tile, and its color represents
        the type of task on that tile at that level.
        Empty circles have no task associated with them.
        The arrows represent dependencies at each level, and
        tasks at each level also implicitly depend on the
        task at the same location in the previous level.}
    \label{fig:QR}
\end{figure}

The tiled QR factorization is based on four basic tasks,
or kernels, as shown in \fig{QR}.
For a matrix consisting of $N\times N$ tiles, $N$ passes,
or levels, are computed, each computing a column and row of the QR
decomposition.
The tasks can be defined in terms of the tuples $(i,j,k)$,
where $i$ and $j$ are the row and column of the tile, respectively,
and $k$ is its level:

\begin{center}
    \begin{tabular}{llll}
        Task & where & depends on task(s) & locks tile(s) \\
        \hline
        \epsfig{file=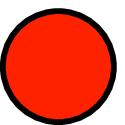,height=9pt} DGEQRF & $i=j=k$ & $(i,j,k-1)$ & $(i,j)$ \\
        \epsfig{file=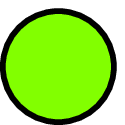,height=9pt} DLARFT & $i=k$, $j>k$ & $(i,j,k-1)$, $(k,k,k)$ & $(i,j)$ \\
        \epsfig{file=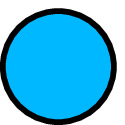,height=9pt} DTSQRF & $i>k$, $j=k$ & $(i,j,k-1)$, $(i-1,j,k)$ & $(i,j)$, $(j,j)$ \\
        \epsfig{file=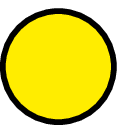,height=9pt} DSSRFT & $i>k$, $j>k$ & $(i,j,k-1)$, $(i-1,j,k)$, $(i,k,k)$ & $(i,j)$ \\
        \hline
    \end{tabular}
\end{center}

\noindent and where the task names are the BLAS-like operation
performed on the given tiles.
Every task depends on the task at the same position and the
previous level, i.e.~the task $(i,j,k)$ always depends on
$(i,j,k-1)$ for $k>1$.
Each task also modifies its own tile $(i,j)$, and the DTSQRF
task additionally modifies the lower triangular part of the $(j,j)$th tile.

Although the tile-based QR decomposition requires only dependencies,
i.e.~no additional conflicts are needed to avoid concurrent access to
the matrix tiles, we still model each tile as a separate resource
in QuickSched such that the scheduler can preferrentially assign
tasks using the same tiles to the same thread.
The resources/tiles are initially assigned to the queues in column-major
order, i.e.~the first $\lfloor n_\mathsf{tiles}/n_\mathsf{queues}\rfloor$
are assigned to the first queue, and so on.

The QR decomposition was computed for a $2048\times 2048$
random matrix using tiles of size $64\times 64$ floats using QuickSched
as described above.
The task costs were initialized to the asymptotic cost of the underlying
operations.
For this matrix, a total of 11\,440 tasks with 21\,824 dependencies,
as well as 1\,024 resources with 21\,856 locks and 11\,408 uses
were generated.

\begin{figure}
    \centerline{\epsfig{file=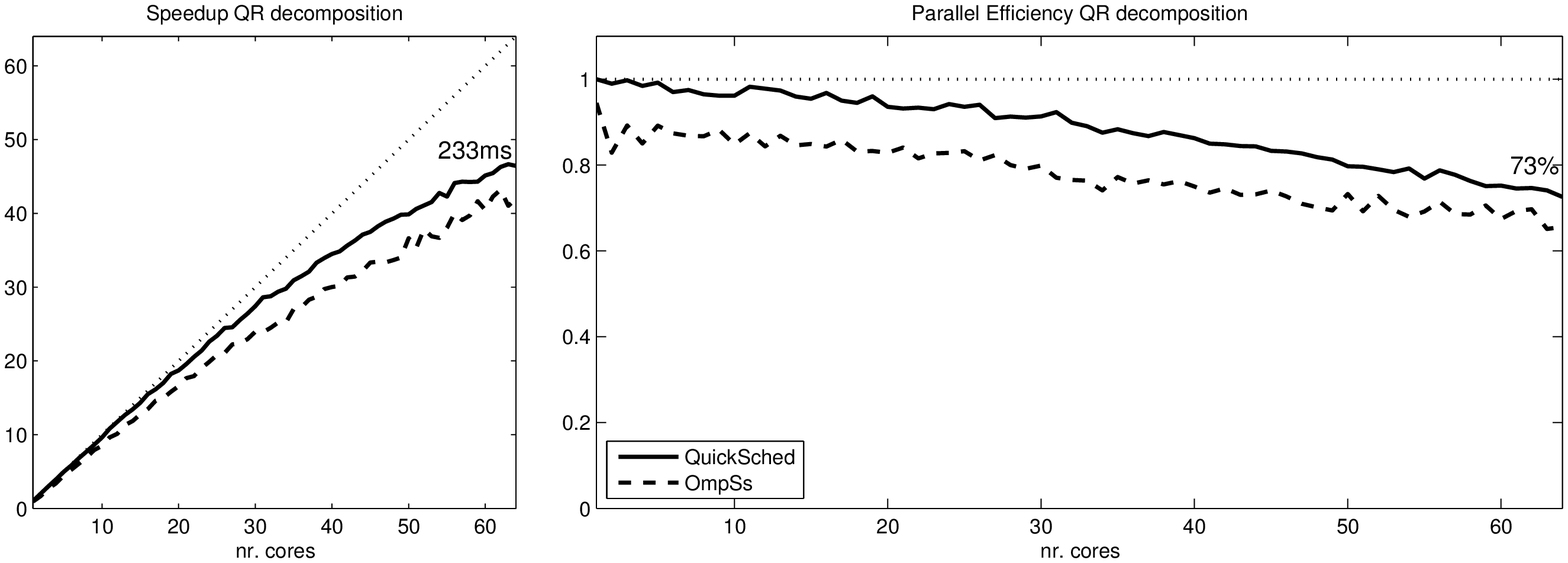,width=\textwidth}}
    \caption{Strong scaling and parallel efficiency of the tiled QR decomposition
        computed over a $2048\times 2048$ matrix with tiles of size
        $64\times 64$.
        The QR decomposition with QuickSched takes 233\,ms,
        achieving 73\% parallel efficiency, over all 64 cores.
        The scaling and efficiency for OmpSs are computed relative to QuickSched.
        }
    \label{fig:QRResults}
\end{figure}

For these tests, {\tt pthread} parallelism and resource re-owning
were used with one queue per core.
The QR decomposition was computed 10 times for each number of
cores, and the average thereof taken for the scaling and
efficiency results in \fig{QRResults}.
The timings are for {\tt qsched\_run}, including the cost of
{\tt qsched\_start}, which does not run in parallel.
Setting up the scheduler, tasks, and resources took, in all
cases, an average of 7.2\,ms, i.e.~at most 3\% of the total
computational cost.

The same decomposition was implemented using OmpSs v.\,1.99.0,
calling the kernels directly using {\tt \#pragma omp task}
annotations with the respective dependencies, and
the runtime parameters
\begin{quote}
  \tt --disable-yield --schedule=socket --cores-per-socket=16 \\--num-sockets=4
\end{quote}
\noindent Several different schedulers and parameterizations
were discussed with the authors of OmpSs and tested, with
the above settings producing the best results.

\begin{figure}
    \centerline{\epsfig{file=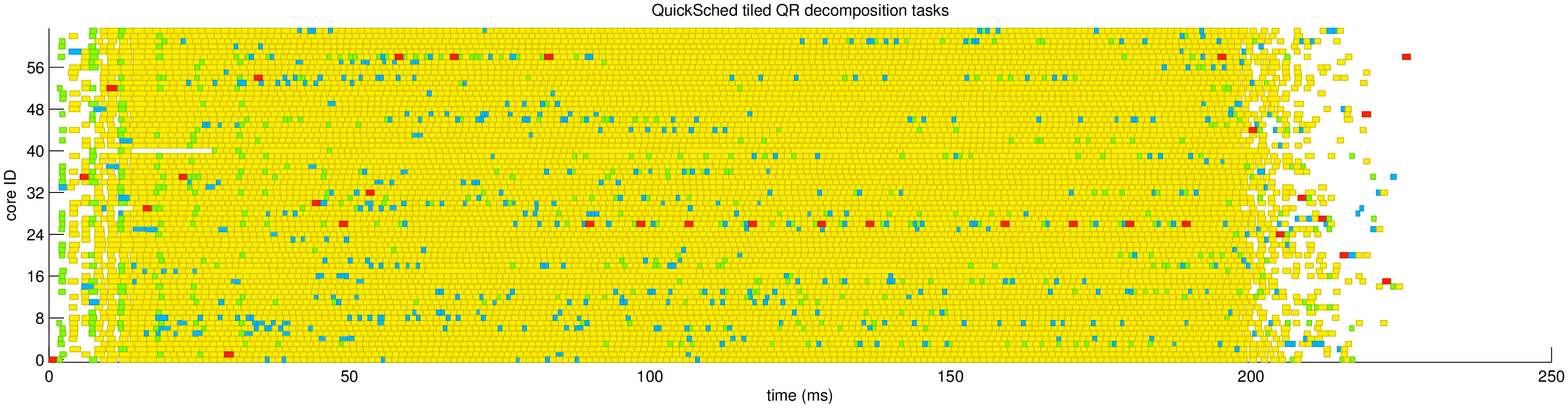,width=\textwidth}}
    \centerline{\epsfig{file=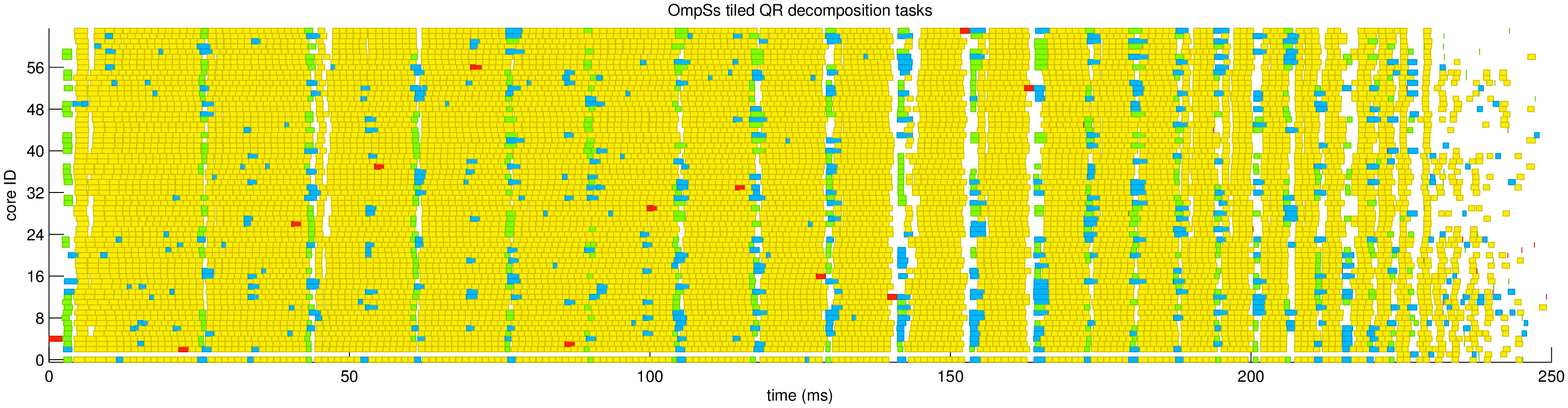,width=\textwidth}}
    \caption{Task scheduling in QuickSched (above) and OmpSs (below)
        for a $2048\times 2048$ matrix on 64 cores.
        The task colors correspond to those in \fig{QR}.}
    \label{fig:QRTasks}
\end{figure}

The scaling and efficiency relative to QuickSched are 
shown in \fig{QRResults}.
The difference in timings is the result of the different
task scheduling policies, as well as a smaller lag between the
individual tasks, as shown in \fig{QRTasks},
which shows the assignment of the different tasks to cores for the
64-core run.
The most visible difference between both schedulers is that
the DGEQRF tasks (in red) are scheduled as soon as they
become available in QuickSched, thus preventing bottlenecks
near the end of the computation.

Since in QuickSched the entire task structure is known explicitly
in advance, the scheduler ``knows'' that the DGEQRF tasks all
lie on the longest critical path and therefore executes them as
soon as possible.
OmpSs does not exploit this knowledge, resulting in the less efficient
scheduling seen in \fig{QRTasks}.

\subsection{Task-Based Barnes-Hut N-Body Solver}

The Barnes-Hut tree-code \citep{ref:Barnes1986}
is an algorithm to approximate the
solution of an $N$-body problem, i.e.~computing all the
pairwise interactions between a set of $N$ particles,
in \oh{N\log N} operations, as opposed to in \oh{N^2} for the
naive direct computation.
The algorithm is based on a recursive octree decomposition:
Starting from a cubic cell containing all the particles,
the cell is recursively bisected along all three spatial dimensions,
resulting in eight sub-cells, until the number of particles
per cell is smaller than some limit $n_\mathsf{max}$.
The particle interactions can then be formulated recursively:
Given a particle and a set of particles in a cell,
if the particle and cell
are sufficiently well separated, the particle-cell interactions
are approximated by interacting the particle with the cell's
center of mass.
If the particle and the cell are too close, and the cell
has sub-cells, i.e.~it contained more than $n_\mathsf{max}$
particles and was split in the recursive octree decomposition,
then the particle is interacted with each of the sub-cells
recursively.
Finally, if the cell is not split, i.e.~it is a leaf cell
in the octree, then the particle is interacted with all
particles in the cell, except for the particle itself if
it is in the same cell.
This operation is performed for each particle, starting
with the root-level cell containing all the particles.

\begin{figure}
    \centerline{\epsfig{file=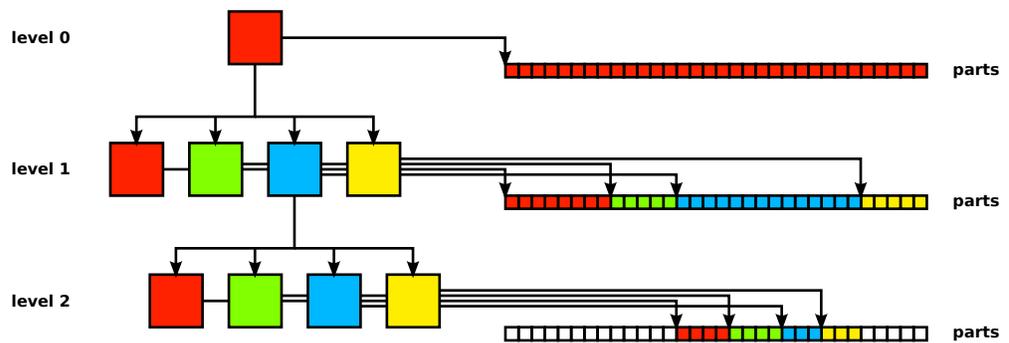,width=0.9\textwidth}}
    \caption{Hierarchical ordering of the particle data structures
    (right) according to their cell (left).
    Each cell has a pointer to the first of its particles (same color
    as cells) in the same global parts array.}
    \label{fig:CellParts}
\end{figure}

In our implementation, the particle data is sorted hierarchically,
following the octree structure.
Unlike in many codes, where the leaves store an array of
pointers to the underlying particles, which are not necessarily
contiguous in memory, the cells, at all
levels, store only a pointer to the first of their own particles,
and the total number of particles.
The current approach, illustrated in \fig{CellParts} is not
only more compact, it also allows a direct and more cache-efficient access
to the list of particles for any inner node of the tree.
The cost of sorting the particles, with a recursive
partitioning similar to QuickSort \citep{ref:Hoare1962},
is in \oh{N\log N}.

The task-based implementation consists of three
types of tasks:
\begin{itemize}
    \item {\em Self}-interactions, in which all particles
        in a single cell interact with all other particles in the
        same cell,
    \item {\em Particle-particle} pair interactions, in which
        all particles in a cell interact with all
        particles in another cell,
    \item {\em Particle-cell} pair interactions, in which
        all particles in one cell are interacted with the
        center of mass of all other cells in the tree.
\end{itemize}

These tasks can be created recursively over the cell hierarchy
as shown in the function {\tt make\_tasks} in \fig{MakeTasks}.
The function is called on the root cell with the root cell
and {\tt NULL} as its two cell parameters.
The function recurses as follows (line numbers refer to \fig{MakeTasks}:
\begin{itemize}
    \item If called with a single, split cell (lines~6--7),
        recurse over all the cell's sub-cells, and all
        pairs of the cell's sub-cells (lines~8--11),
    \item If called with a single unsplit cell (line~13),
        create a self-interaction task (line~14) as well as a particle-cell
        task on that cell (line~18),
    \item If called with two neighbouring cells and both cells
        are split (line~22),
        recurse over all pairs of sub-cells spanning
        both cells (lines~24--26), and
    \item If called with two neighbouring cells
        and at least one of the cells is not split, create
        a particle-particle pair task over both cells (line~29),
    \item If called with two non-neighbouring cells,
        do nothing, as these interactions
        will be computed by the particle-cell task.
\end{itemize}
\noindent Every interaction task additionally locks
the cells on which it operates (lines~17, 20, and 32--33).
In order to prevent generating
a large number of very small tasks, the task generation only recurses
if the cells contain more than a minimum number $n_\mathsf{task}$
of particles each (lines~7 and~23).

As shown in \fig{BHTasks}, the particle-self and particle-particle pair
interaction tasks are implemented
by computing the interactions between all particle pairs spanning
both cells in a double for-loop (lines~9--11 and~22-24 therein).
Some extra logic (lines~2--7 and~15--20) is added to deal with
split cells that did not contain enough particles to warrant the
generation of finer tasks.
The particle-cell interactions for each leaf node are computed by
traversing the tree recursively starting from the root node and:
\begin{itemize}
  \item If called with a node that is a hierarchical parent of
    the leaf node, or with a node that is a direct neighbour of
    a hierarchical parent of the leaf node, recurse over the
    node's sub-cells (lines~29--32),
  \item Otherwise, compute the interaction between the leaf node's
    center of mass and all the particles in the leaf node (lines~33--35).
\end{itemize}

This task decomposition differs from the traditional tree-walk
in the Barnes-Hut algorithm in that the particle-cell interactions
are grouped per leaf, with each leaf doing its own tree walk,
as opposed to doing a tree walk for each individual particle.
This approach was chosen to maximize the memory locality
of the particle-cell calculation, as the particles in the leaf,
which are traversed for each particle-cell interaction, are
contiguous in memory, and are thus more likely to remain in the
lowest-level cache during the entire tree-walk.

This Barnes-Hut tree-code was used to approximate the gravitational
N-Body problem for 1\,000\,000 particles with uniformly random coordinates
in $[0,1]^3$.
The parameters $n_\mathsf{max}=100$ and $n_\mathsf{task}=5000$
were used to generate the tasks.
Using the above scheme generated 97\,553 tasks, of which
512 self-interaction tasks, 5\,068 particle-particle interaction
task, and 32\,768 particle-cell interaction tasks.
A total of 43\,416 locks on 37\,449 resources were generated.
Setting up the tasks took, on average XXX\,ms, i.e.~at most
XXX\% of the total computation time.
Storing the tasks, resources, and dependencies required XXX\,MB,
which is only XX\% of the XXX\,MB required to store the particle
data.

\begin{figure}
    \centerline{\epsfig{file=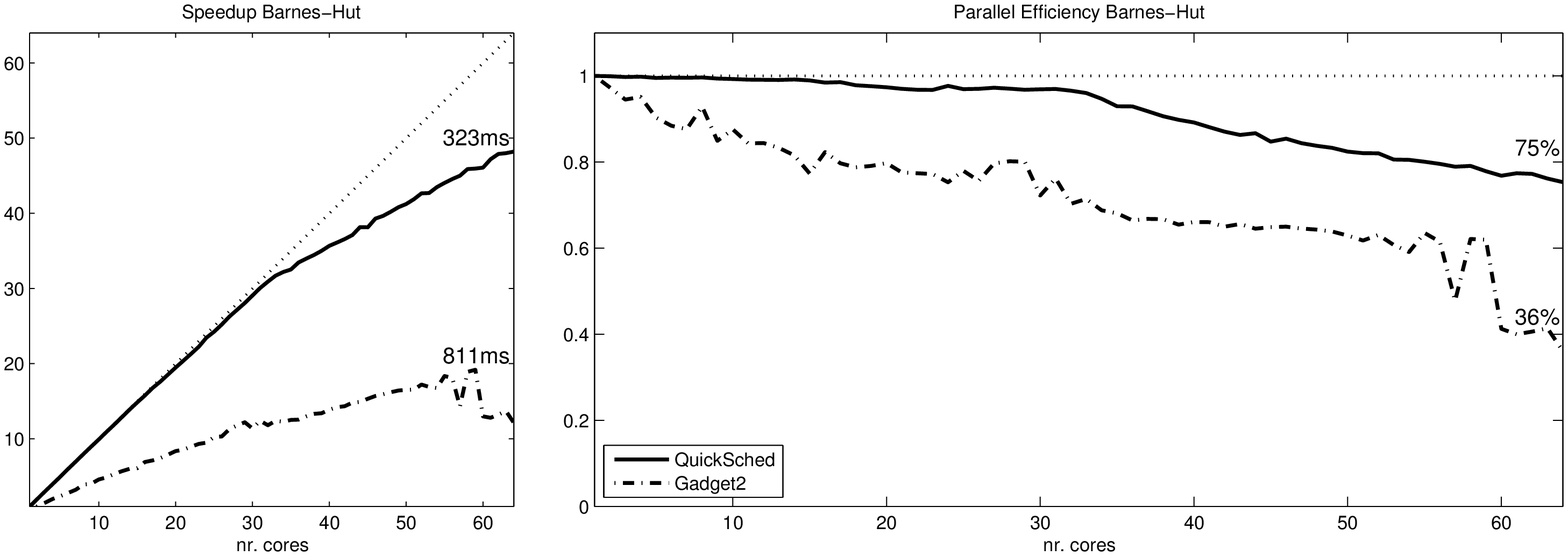,width=\textwidth}}
    \caption{Strong scaling and parallel efficiency of the Barnes-Hut tree-code
        computed over 1\,000\,000 particles.
        Solving the N-Body problem takes 323\,ms, achieving 75\% parallel
        efficiency over all 64 cores.
        For comparison, timings are shown for the same computation using
        the popular astrophysics code Gadget-2.
        The scaling for Gadget-2 (left) is shown relative to the performance of
        QuickSched, whereas the parallel efficiency (right) is computed relative
        to Gadget-2 on a single core.
        }
    \label{fig:BHResults}
\end{figure}

For these tests, {\tt pthread}s parallelism was used and resource
re-owning was switched off.
Resource ownership was attributed by dividing the global
{\tt parts} array by the number of queues and assigning each cell's
{\tt res} to the fraction of the {\tt parts} array to which
the first of its own {\tt parts} belong.
The interactions were computed 10 times for each number of
cores, and the average thereof taken for the scaling and
efficiency results in \fig{BHResults}.
The timings are for {\tt qsched\_run}, including the cost of
{\tt qsched\_start}, which does not run in parallel.
Setting up the scheduler, tasks, and resources took, in all
cases, an average of 51.3\,ms.

For comparison, the same computations were run using the popular
astrophysics simulation software Gadget-2 \citep{ref:Springel2005},
using a traditional Barnes-Hut implementation based on octrees
and distributed-memory parallelism based on domain decompositions
and MPI \citep{ref:Snir1998}.
To achieve the same accuracy, an opening angle of 0.5 was used.
On a single core, the task-based tree traversal is already 1.9$\times$
faster than Gadget-2, due to the cache efficiency of the task-based
computations, which, by design, maximize the amount of computation
per memory access.
At 59 cores, where Gadget-2 performs best, the task-based tree traversal is
2.51$\times$ faster, and at the full 64 cores it is 4$\times$ faster,
due to the better strong scaling of the task-based approach as opposed
to the MPI-based parallelism in Gadget-2.

\begin{figure}
    \centerline{\epsfig{file=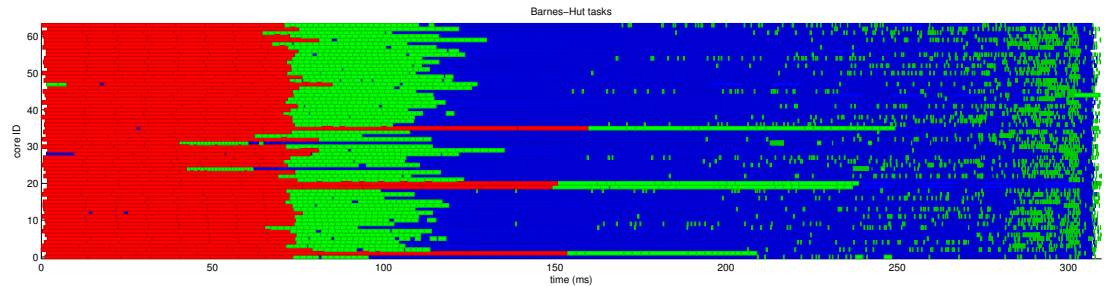,width=\textwidth}}
    \caption{Task scheduling of the Barnes-Hut tree-code on 64 cores.
      The red tasks correspond to particle self-interactions, the green
      tasks to the particle-particle pair interactions, and the blue
      tasks to the particle-cell interactions.}
    \label{fig:BHTasksPlot}
\end{figure}

\begin{figure}
    \centerline{\epsfig{file=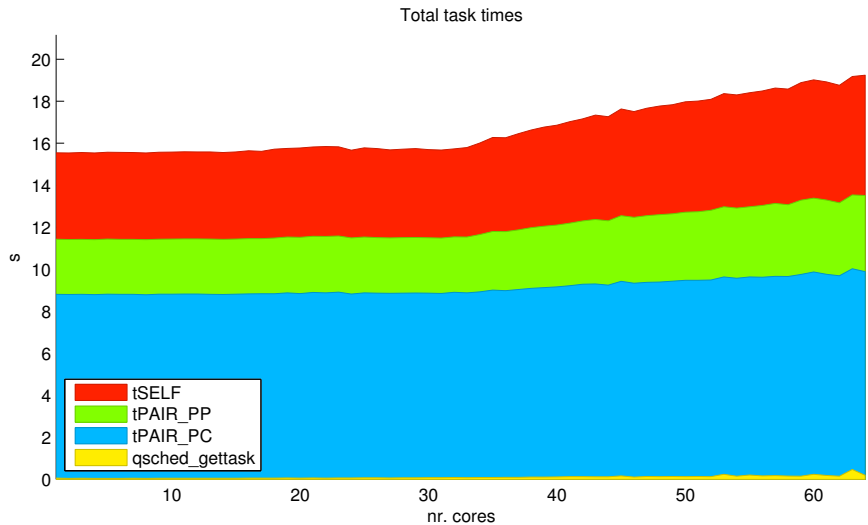,width=0.8\textwidth}}
    \caption{Accumulated cost of each task type and of the overheads
        associated with {\tt qsched\_gettask}, summed over all cores.
        As of 32 cores, the cost of both pair interaction task
        types grow by up to 30\%.
        The cost of the particle-cell interactions, which entail significantly more
        computation per memory access, grow only by at most 10\%.
        The scheduler overheads, i.e.~{\tt qsched\_gettask},
        make up less than 1\% of the total time.}
    \label{fig:BHTimes}
\end{figure}

Unlike the QR decomposition, the results scale well only to
32 cores, achieving 90\% parallel efficiency, and then
level off for increasing numbers of cores.
This, however, is not a problem of the task-based parallel
algorithm, or of QuickSched, but of the memory bandwidth
of the underlying hardware.
\fig{BHTimes} shows the accumulated cost of each task type and of 
QuickSched over the number of cores.
At 64 cores, the scheduler overheads account for only $\sim 1$\% of
the total computational cost, whereas,
as of 32 cores, the cost of both pair types grow by up to
40\%.
This is due to the cache hierarchy of the AMD Opteron 6376 in which
pairs of cores share a comon 2\,MB L2 cache.
When using half the cores or less, each core has its L@ cache to
itself, whereas beyond 32 cores they are shared, resulting in more
frequent cache misses.
This cen be seen when comparing the costs of the particle-particle
interaction and particle-cell interaction tasks: while the former grow by
roughly 30\%, the latter grow by only 10\% as they do much more
computation per memory access.

\section{Discussion and Conclusions}

The task scheduler described in the previous sections, QuickSched,
differs from existing task-based programming schemes
in a number of ways.
The most obvious such difference is the addition of {\em conflicts},
modeled using exclusive locks on hierarchical resources.
This extension to the standard dependencies-only model
of task-based parallelism allows for more complex task relations,
such as in the Barnes-Hut tree-code described earlier.

Another significant difference is that the tasks, their
dependencies, and their conflicts must be described
{\em explicitly} before the parallel computation starts.
This as opposed to implicit dependencies generated
by task spawning, e.g.~as in Cilk, or to extracting the
dependencies from the task parameters, e.g.~in QUARK or OmpSs.
Explicitly defining dependencies has the advantage that 
more elaborate dependency structures can be generated.
Furthermore, knowing the structure of the entire task
graph from the start of the computation provides valuable
information when scheduling the tasks, e.g.~using the 
critical path along the dependencies to compute the
task weight.

Finally, as opposed to the most other task-based
programming environments which rely on compiler extensions
and/or code pre-processors, QuickSched operates as a regular
C-language library, based on standard parallel functionality
provided by OpenMP and/or {\tt pthreads}.
This ensures a maximum amount of portability on existing
and future architectures.
The interfaces are also kept as simple
as possible in order to reduce the burden on the programmer
when implementing task-based codes.

The QuickSched library itself is remarkably simple, consisting of
less than 3\,000 lines of code, including comments.
Both examples, which are distributed with QuickSched,
require less than 1\,000 lines of code each.
For a more complex, large-scale example of a task-based computation
based on the same algorithms, we refer to \cite{ref:Gonnet2014},
for which the scheduler was originally designed, and from
which QuickSched was back-ported.

In both examples, QuickSched performs extremely well, even
on a large number of shared-memory cores.
This performance is due, on the one hand, to the
division of labor between the scheduler and the queues,
and on the other hand due to the simple yet efficient
algorithms for task selection and resource locking.
The task weighting based on the length of the critical
path of the dependencies delivers, in the examples shown,
good parallel efficiency.

There are several possible improvements to QuickSched which
have not been addressed in this paper.
The most obvious of which are the following:
\begin{itemize}
    \item {\em Priorities}: The current implementation of
        QuickSched does not take the resource locks into
        account when selecting tasks in the queues, e.g.~it
        may be advantageous, in some settings, to avoid tasks
        which are involved in too many potential conflicts
        and would therefore restrict the maximum degree of
        parallelism when scheduled.
    \item {\em Work-stealing}: During work-stealing, the
        queues are probed in a random order although
        the total relative cost of the tasks in the queue,
        as well as the length of their critical paths are
        known,
    \item {\em Costs}: The size of the resources used by
        a task are currently not taken into account when
        assigning it to the queues in {\tt qsched\_enqueue},
        or when approximating the cost of a task.
\end{itemize}

QuickSched is distributed under the GNU Lesser General Public Licence
v\,3.0 and is available for download via
\url{https://sourceforge.net/projects/quicksched/files/}.

\section*{Acknowledgments}
The authors would like to thank Tom Theuns and Richard Bowers of the
Institute for Computational Cosmology at Durham University for the
helpful discussions.
This work was supported by a Durham University Seedcorn Grant
number 21.12.080130 from
which the hardware used in the experiments was purchased.

\bibliography{quicksched}

\appendix
\section{User Interface}

This section describes the QuickSched interface functions and how they
are called.

As mentioned previously, the {\tt qsched} object is the main
interface to the task scheduler.
As such, it provides functionality for task and resource
creation, for assigning resources to tasks, either as locks
or uses, and for assigning dependencies between tasks.
The tasks and resources themselves are opaque to the
user, and handles of the types {\tt qsched\_task\_t}
and {\tt qsched\_res\_t} are used instead.

The main functions for setting up the scheduler are:
\begin{itemize}
    \item {\tt void qsched\_init(struct qsched *s, int nr\_queues, int flags)} \\
        Initializes a {\tt qsched} object with the given number of queues.
        The {\tt flags} parameter can be set to any bitwise or combination
        of {\tt qsched\_flag\_none},
        {\tt qsched\_flag\_yield}, and {\tt qsched\_flag\_pthread},
        which are described further below.
        This function must be called before any of the other
        functions are used.
        \vspace{1mm}
    \item {\tt void qsched\_free(struct qsched *s)} \\
        Releases all the memory and other resources allocated by the
        given {\tt qsched} object.
        After this function has been called, the {\tt qsched} will
        need to be re-initialized for reuse.
        \vspace{1mm}
    \item {\tt void qsched\_reset(struct qsched *s)} \\
        Clears the tasks and resources in the given {\tt qsched},
        but does not release the allocated memory or change
        the number of queues.
        \vspace{1mm}
    \item {\tt qsched\_task\_t qsched\_addtask(struct qsched *s, int type, unsigned int flags, void *data, int data\_size, int cost)} \\
        Creates a new task within the given {\tt qsched} and returns
        its handle.
        The {\tt type} and {\tt data} field are copied into the {\tt qsched}
        and passed to the execution function when the {\tt qsched} is run.
        The parameter flags is either {\tt task\_flag\_none} or
        {\tt task\_flag\_virtual}.
        Tasks marked as virtual do not have any action associated with them,
        e.g. they are used only to group or otherwise dependencies, and
        are not passed to the execution function in {\tt qsched\_run}.
        \vspace{1mm}
    \item {\tt qsched\_res\_t qsched\_addres(struct qsched *s, int owner, qsched\_res\_t parent)} \\
        Creates a new resource within the given {\tt qsched} and returns
        its handle.
        The owner field is the initial queue ID to which this resource
        should be assigned, or {\tt qsched\_owner\_none}.
        The {\tt parent} field is the handle of the hierarchical parent of
        the new resource or {\tt qsched\_res\_none} if the resource
        has no hierarchical parent.
        \vspace{1mm}
    \item {\tt void qsched\_addlock(struct qsched *s, qsched\_task\_t t, qsched\_res\_t res)} \\
        Append the resource {\tt res} to the task {\tt t}'s list of
        locks.
        The task {\tt t} will then conflict with any other task that
        also locks the resource {\tt res}, its hierarchical parents, or
        any resource hierarchically below it.
        \vspace{1mm}
    \item {\tt void qsched\_adduse(struct qsched *s, qsched\_task\_t t, qsched\_res\_t res)} \\
        Similar to {\tt qsched\_addlock}, yet the resource is only used and
        is not part of a conflict.
        This information is used when assigning tasks to specific queues.
        \vspace{1mm}
    \item {\tt void qsched\_addunlock(struct qsched *s, qsched\_task\_t ta, qsched\_task\_t tb)} \\
        Appends the task {\tt tb} to the list of tasks that the task {\tt ta}
        unlocks.
        The task {\tt tb} then depends on the task {\tt ta}.
        \vspace{1mm}
    \item {\tt void qsched\_run(struct qsched *s, int nr\_threads, qsched\_funtype fun)} \\
        Executes the tasks in the given {\tt qsched} using {\tt nr\_threads}
        threads via the execution function {\tt fun}, as described in the
        previous section.
        Once a {\tt qsched} has been initialized and filled with
        tasks and resources, it can be run more than once.
        \vspace{1mm}
\end{itemize}

The library can be compiled to use OpenMP and/or the
{\tt pthreads} library.
OpenMP is the default, but calling {\tt qsched\_init} with
either the {\tt qsched\_flag\_yield}
or the {\tt qsched\_flag\_pthread} switches to using {\tt pthreads},
if available, for the parallel loop.

OpenMP has the advantage of being available for most compilers
and also potentially providing some extra platform-specific
scheduling features, e.g.~optimal thread location and/or affinity,
and of integrating seamlessly with other parallel parts of the
user application.
The disadvantage of using OpenMP is that it does not provide
any mechanism for yielding a thread if no tasks are available,
i.e. the main loop in {\tt qsched\_gettask}, described in the
previous section, will spin until a task becomes available.
This may be a problem if other parts of the user application
are running concurrently in the background outside of QuickSched.
Calling {\tt qsched\_init} with the {\tt qsched\_flag\_yield}
forces the use of {\tt pthreads} and uses conditional variables
to wait for a new task to be enqueued if obtaining a task
from any of the queues fails.
This relinquishes the waiting computational thread for other
processes.

\section{Tiled QR Implementation}

\begin{figure}
\begin{center}\begin{minipage}{0.9\textwidth}
    \begin{lstlisting}[basicstyle=\scriptsize\tt]
enum {tDGEQRF, tDLARFT, tDTSQRF, tDSSRFT};
void make_tasks(struct qsched *s, int m, int n) {
  int i, j, k, data[3];
  qsched_task_t tid[m * n], tid_new;
  qsched_res_t rid[m * n];
  for (k = 0; k < m * n; k++) {
    tid[k] = qsched_task_none;
    rid[k] = qsched_addres(s, qsched_res_none);
  }
  for (k = 0; k < m && k < n; k++) {
    /* DGEQRF task at (k,k). */
    data[0] = k; data[1] = k; data[2] = k;
    tid_new = qsched_addtask(s, tDGEQRF, qsched_flags_none, data,
                             sizeof(int) * 3, 2);
    qsched_addlock(s, tid_new, rid[k * m + k])
    if (tid[k * m + k] != qsched_task_none)
      qsched_addunlock(s, tid[k * m + k], tid_new);
    tid[k * m + k] = tid_new;
    for (j = k + 1; j < n; j++) {
      /* DLARFT task at (k,j). */
      data[0] = k; data[1] = j; data[2] = k;
      tid_new = qsched_addtask(s, tDLARFT, qsched_flags_none, data,
                               sizeof(int) * 3, 3);
      qsched_addlock(s, tid_new, rid[j * m + k])
      qsched_addunlock(s, tid[k * m + k], tid_new);
      if (tid[j * m + k] != qsched_task_none)
        qsched_addunlock(s, tid[j * m + k], tid_new);
      tid[j * m + k] = tid_new;
    }
    for (i = k + 1; i < m; i++) {
      /* DTSQRF task at (i,k). */
      data[0] = i; data[1] = k; data[2] = k;
      tid_new = qsched_addtask(s, tDTSQRF, qsched_flags_none, data,
                               sizeof(int) * 3, 3);
      qsched_addlock(s, tid_new, rid[k * m + i])
      qsched_addlock(s, tid_new, rid[k * m + k])
      qsched_addunlock(s, tid[k * m + k], tid_new);
      if (tid[k * m + i] != qsched_task_none)
        qsched_addunlock(s, tid[k * m + i], tid_new);
      tid[k * m + i] = tid_new;
      for (j = k + 1; j < n; j++) {
        /* DSSRFT task at (i,j). */
        data[0] = i; data[1] = j; data[2] = k;
        tid_new = qsched_addtask(s, tDSSRFT, qsched_flags_none, data,
                                 sizeof(int) * 3, 5);
        qsched_addlock(s, tid_new, rid[j * m + i])
        qsched_addunlock(s, tid[j * m + k], tid_new);
        qsched_addunlock(s, tid[k * m + i], tid_new);
        if (tid[j * m + i] != qsched_task_none)
          qsched_addunlock(s, tid[j * m + i], tid_new);
        tid[j * m + i] = tid_new;
      }
    }
  }
}
    \end{lstlisting}
\end{minipage}\end{center}
\caption{Example code to generate the tasks for the tiled
    QR decomposition.}
\label{fig:CodeQR}
\end{figure}

Setting up the dependencies and locks for a matrix of
$m\times n$ tiles is implemented as shown in \fig{CodeQR},
where the $m\times n$ matrix {\tt tid} stores the handles
of the last task at position $(i,j)$ and is initialized with
empty tasks (line~7).
Similarly, {\tt rid} stores the handles of the resources for each
tile of the matrix, which are allocated in line~8.

The following loops mirror the task generation described in
Algorithm~2 of \citep{ref:Buttari2009}.
For each level {\tt k} (line~10), a DGEQRF task is created
for tile $(k,k)$ (lines~13--14).
A lock is added for the newly created task on the
resource associated with the $(k,k)$th tile (line~15).
If a task exists at that position at the previous level
(line~16), a dependency is added between the old task and
the new (line~17), and the new task is stored in {\tt tid}
(line~18).
The remaining tasks are generated in the same way, with
their respective locks and dependencies.

The execution function for these tasks simply calls the appropriate
kernels on the matrix tiles given by the task data:
\begin{center}\begin{minipage}{0.9\textwidth}
    \begin{lstlisting}
void exec_fun(int type, void *data) {
  int *idata = (int *)data;
  int i = idata[0], j = idata[1], k = idata[2];
  switch (type) {
    case tDGEQRF:
      DGEQRF(A[i, j], ...);
      break;
    case tDLARFT:
      DLARFT(A[i, j], A[i, i], ...);
      break;
    case tDTSQRF:
      DTSQRF(A[i, j], A[j, j], ...);
      break;
    case tDSSRFT:
      DSSRFT(A[i, j], A[i, k], A[k, j], ...);
      break;
    default:
      error("Unknown task type.");
  }
}
    \end{lstlisting}
\end{minipage}\end{center}
\noindent where {\tt A} is the matrix over which the QR
decomposition is executed.

\section{Barnes-Hut N-body Solver Implementation}

The cells themselves are implemented using the following 
data structure:
\begin{center}\begin{minipage}{0.9\textwidth}
    \begin{lstlisting}
struct cell {
  double loc[3], h[3], com[3], mass;
  int split, count;
  struct part *parts;
  struct cell *progeny[8];
  qsched_res_t res;
  qsched_task_t task_com;
};
    \end{lstlisting}
\end{minipage}\end{center}
\noindent where {\tt loc} and {\tt h} are the location
and size of the cell, respectively.
The {\tt com} and {\tt mass} fields represent the cell's
center of mass, which will be used in the particle-cell interactions.
The {\tt res} filed is the hierarchical resource representing
the cell's particles, and it is the parent resource of the cell
progeny's {\tt res}.
Similarly, the {\tt task\_com} is a task handle to
compute the center of mass of the cell's particles, and 
it depends on the {\tt task\_com} of all the progeny if
the cell is split.
{\tt parts} is a pointer to an array of {\tt count} 
particle structures, which contain all the particle
data of the form:
\begin{center}\begin{minipage}{0.9\textwidth}
    \begin{lstlisting}
struct part {
  double x[3], a[3], mass;
  int id;
};
    \end{lstlisting}
\end{minipage}\end{center}
\noindent i.e.~the particle position, acceleration, mass,
and ID, respectively.

\begin{figure}
\begin{center}\begin{minipage}{0.9\textwidth}
    \begin{lstlisting}[basicstyle=\scriptsize\tt]
void comp_self(struct cell *c) {
  if (c->split) {
    for (int j = 0; j < 8; j++) {
      comp_self(c->progeny[j]);
      for (int k = j + 1; k < 8; k++)
        comp_pair(c->progeny[j], c->progeny[k]);
    }
  } else {
    for (int j = 0; j < c->count; j++)
      for (int k = j + 1; k < c->count; k++)
        interact c->parts[j] and c->parts[k].
  }
}

void comp_pair(struct cell *ci, struct cell *cj) {
  if (ci and cj are not neighbours)
    return;
  if (ci->split && cj->split) {
    for (int j = 0; j < 8; j++)
      for (int k = 0; k < 8; k++)
        comp_pair(ci->progeny[j], cj->progeny[k]);
  } else {
    for (int j = 0; j < ci->count; j++)
      for (int k = 0; k < cj->count; k++)
        interact ci->parts[j] and cj->parts[k].
  }
}

void comp_pair_cp(struct cell *leaf, struct cell *c) {
  if (c is a parent of leaf ||
      c is a neighbour of a parent of leaf) {
    for (int k = 0; k < 8; k++)
     comp_pair_cp(leaf, c->progeny[k]);
  } else if (leaf and c are not direct neighbours) {
    for (int k = 0; k < leaf->count; k++)
      interact leaf->parts[k] and c center of mass.
  }
}
    \end{lstlisting}
\end{minipage}\end{center}
    \caption{Task functions for the Barnes-Hut tree-code.}
    \label{fig:BHTasks}
\end{figure}

\begin{figure}
\begin{center}\begin{minipage}{0.9\textwidth}
    \begin{lstlisting}[basicstyle=\scriptsize\tt]
enum { tSELF, tPAIR_PP, tPAIR_PC };
void make_tasks(struct qsched *s, struct cell *ci, struct cell *cj) {
  int j, k;
  qsched_task_t tid;
  struct cell *data[2];
  if (cj == NULL) {
    if (ci->split && ci->count > n_task) {
      for (j = 0; j < 8; j++) {
        make_tasks(s, ci->progeny[j], NULL);
        for (k = j + 1; k < 8; k++)
          make_tasks(s, ci->progeny[j], ci->progeny[k]);
      }
    } else {
      tid = qsched_addtask(s, tSELF, qsched_flags_none, &ci,
                           sizeof(struct cell *),
                           ci->count * ci->count);
      qsched_addlock(s, tid, ci->res);
      tid = qsched_addtask(s, tPAIR_PC, qsched_flags_none, &ci,
                           sizeof(struct cell *), ci->count);
      qsched_addlock(s, tid, ci->res);
    }
  } else if (ci->split && cj->split &&
             ci->count * cj->count > n_task * n_task) {
    for (j = 0; j < 8; j++)
      for (k = 0; k < 8; k++)
        make_tasks(s, ci->progeny[j], cj->progeny[k]);
  } else {
    data[0] = ci; data[1] = cj;
    tid = qsched_addtask(s, tPAIR_PP, qsched_flags_none, data,
                         sizeof(struct cell *) * 2,
                         ci->count * cj->count);
    qsched_addlock(s, tid, ci->res);
    qsched_addlock(s, tid, cj->res);
  }
}
    \end{lstlisting}
\end{minipage}\end{center}
    \caption{C-like pseudo-code for recursive task creation
        for the Barnes-Hut tree-code.}
    \label{fig:MakeTasks}
\end{figure}

The functions for the task themselves are relatively
straight-forward and shown in \fig{BHTasks}, and the
execution function can be written as:
\begin{center}\begin{minipage}{0.9\textwidth}
    \begin{lstlisting}
void exec_fun(int type, void *data) {
  struct cell **cells = (struct cell **)data;
  switch (type) {
    case tSELF:
      comp_self(cells[0]);
      break;
    case tPAIR_PP:
      comp_pair(cells[0], cells[1]);
      break;
    case tPAIR_PC:
      comp_pair_pc(cells[0], root);
      break;
    case tCOM:
      comp_com(cells[0]);
      break;
    default:
      error("Unknown task type.");
  }
}
    \end{lstlisting}
\end{minipage}\end{center}

\end{document}